\documentclass[aps,prl,twocolumn,reprint,amsmath,superscriptaddress]{revtex4}

\usepackage{graphicx}
 
\begin{document}

\title{Exotic Superfluid States of Lattice Fermions in Elongated Traps}

\author{D.-H. Kim}
\affiliation{Department of Applied Physics, Aalto University, P.O. Box 15100, 00076 AALTO, Finland}

\author{J. J. Kinnunen}
\affiliation{Department of Applied Physics, Aalto University, P.O. Box 15100, 00076 AALTO, Finland}

\author{J.-P. Martikainen}
\affiliation{NORDITA, 106 91 Stockholm, Sweden}

\author{P. T\"{o}rm\"{a}}
\email{paivi.torma@hut.fi}
\affiliation{Department of Applied Physics, Aalto University, P.O. Box 15100, 00076 AALTO, Finland}


\begin{abstract}
We present real-space dynamical mean-field theory calculations 
for attractively interacting fermions in three-dimensional lattices 
with elongated traps. The critical polarization is found to be 0.8, 
regardless of the trap elongation. Below the critical polarization, 
we find unconventional superfluid structures where the polarized 
superfluid and Fulde-Ferrell-Larkin-Ovchinnikov-type states
emerge across the entire core region. 
\end{abstract}


\maketitle

The nature of pairing in spin-polarized fermion systems is 
a fundamental problem in many areas of physics, including superconductors 
in a strong magnetic field, neutron-proton pairing in nuclear matter, 
and color superconductivity in high density QCD
\cite{Casalbuoni2004,Radovan2003,Kenzelmann2008}.
Non-BCS pairing mechanisms have been proposed for spin-polarized fermions. 
The Fulde-Ferrell-Larkin-Ovchinnikov (FFLO) states exhibit finite-momentum 
pairing, causing spatially oscillating pair potentials~\cite{FF,LO}.
In the polarized superfluid states, so-called Sarma or breached-pair (BP)
states, zero-momentum pairing occurs, having excess unpaired 
particles~\cite{Sarma1963,Liu2003,Sheehy2006,Parish2007,Pilati2008}.
In trapped ultracold Fermi gases~\cite{Bloch2008,Giorgini2008}, 
recent experiments with spin imbalance observed
the superfluidity and the transition to the normal state 
with increasing polarization
\cite{Zwierlein2006,Partridge2006,Shin2006,Nascimbene2009}.
These systems offer an unprecedented access 
to the search for exotic superfluid states.
In this Letter, we present the first large-scale calculations beyond 
mean-field theory, explicitly considering the trap geometry. 
We characterize the critical polarization of the transition and 
show evidence for exotic superfluidity. 

The FFLO and polarized superfluid states have been originally proposed for 
translationally invariant systems. The FFLO state, in particular, 
has been suggested to be stabilized by reduced dimensionality. 
Therefore, it is important to incorporate broken translational symmetry
caused by the presence of a trap in ultracold gases.
Highly elongated traps were used in the experiments with spin-imbalanced $^6$Li gases
\cite{Zwierlein2006,Partridge2006,Shin2006,Nascimbene2009,Liao2010}.
Concerning previous mean-field theory calculations with explicit trap 
confinement~\cite{Mizushima2005}, 
the FFLO-type oscillations have been expected to occur only at the narrow 
edges of the superfluid core in a spherical trap~\cite{Kinnunen2006}. 
Quantum fluctuations are largely neglected in these mean-field calculations, 
and a full-scale {\it ab initio} approach remains elusive so far 
in three dimensions. 
Here we address this issue via an extension of dynamical mean-field theory 
considering both the full local quantum fluctuations and the trap effects.     

We study the problem of a trapped lattice fermion system 
at zero temperature ($T=0$) in three dimensions (3D) by using 
a \textit{real-space} dynamical mean-field theory 
(DMFT)~\cite{DMFTreview,Helmes2008,Snoek2008,Koga_arxiv}. 
With various trap aspect ratios examined for elongated traps, 
we find the critical polarization near $0.8$ insensitive to 
the trap aspect ratios. In a wide range of polarizations, 
we observe that polarization and finite pair potential
coexist at the core region, indicating the polarized superfluid core, 
surrounded by the normal unpaired particles.
In highly elongated traps, as we approach the critical polarization, 
we find that the FFLO-type states emerge with spatially 
oscillating pair potential across the trap center. 
For comparison, we present also 
Bogoliubov--de Gennes (BdG) mean-field calculations.

\begin{figure}
\includegraphics[width=0.47\textwidth]{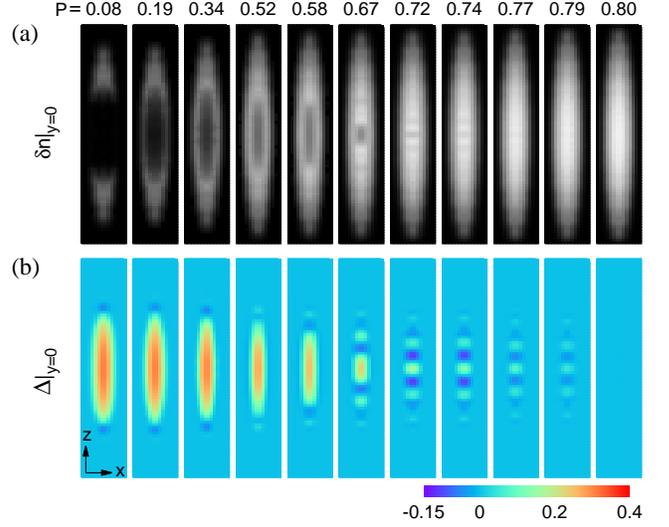} 
\caption{
Cross-sectional visualization of the particle density and pair potential 
structure in the elongated trap with the aspect ratio $\alpha=7.5$ 
for selected polarizations $P \le 0.8$.
(a) The density difference $\delta n \equiv n_\uparrow - n_\downarrow$
is plotted in the $y=0$ plane. The lighter (darker) gray indicates 
the larger (smaller) value of $\delta n$. Black defines $\delta n=0$.
(b) The pair potential $\Delta(x,y=0,z)$ shows
the evolution of the superfluid core structure as a function of $P$.
The FFLO-like oscillations are observed at high $P$'s but below 
$P=0.8$ where $\Delta$ vanishes. 
}
\label{fig1}
\end{figure}

Our real-space DMFT solves the attractive Hubbard model with an 
\textit{anisotropic} trapping potential at $T=0$,
\begin{equation*}
\mathcal{H} = -t \sum_{\langle ij \rangle \sigma} 
c^\dagger_{i\sigma} c_{j\sigma}
- U \sum_i n_{i\uparrow} n_{i\downarrow} 
+ \sum_{i\sigma} (V_i-\mu_\sigma) n_{i\sigma} ,
\end{equation*}
where $c^\dagger_{i\sigma}$($c_{i\sigma}$) creates (annihilates) 
a fermion with spin $\sigma$ at site $i$, the density operator 
$n_{i\sigma} = c^\dagger_{i\sigma} c_{i\sigma}$, 
and $\mu_\sigma$ denotes the chemical potential.
The hopping $t$ between neighboring sites 
$\langle ij \rangle$ is set to unity. The onsite interaction 
$U$ is chosen as the unitarity value for the cubic lattice,
$U \simeq 7.915$~\cite{Burovski2006}. 
The  trapping potential is given as
$V_i=V_0 [{x^\prime}^2+{y^\prime}^2+(z^\prime/\alpha)^2]$
where $V_0$ and $\alpha$ are 
the trap strength and aspect ratio. In the 3D 
lattice with size $L_x \times L_y \times L_z$,
the coordinates of sites are assigned as $\xi^\prime = \xi-\frac{1}{2}$, where 
$\xi \in \{x,y,z\}$, and $\xi = -\frac{L_\xi}{2}, \ldots, \frac{L_\xi}{2}-1$.

The real-space DMFT is constructed for inhomogeneous systems by 
considering local, yet site dependent, self-energy terms, which explicitly 
include trap effects beyond the local density approximation (LDA). 
This approximation in the DMFT allows s-wave pairing~\cite{DMFTreview}. 
The impurity problem now becomes site-dependent 
and must be solved for each site with an efficient solver. 
We employ the exact diagonalization (ED) method to solve 
a $7$-orbital impurity Hamiltonian with a superconducting 
bath~\cite{DMFTreview} which provides a good convergence
(cf.~\cite{Barnea2008}) in our reliability tests.
There are two advantages in choosing the ED method:  
it is easy to consider the generalized Anderson model  
required to describe the superfluid phases;
it efficiently finds the ground state. 
The possible source of error is the effective bath 
of a limited number of orbitals, causing spiky spectral functions. 

We consider harmonic trapping potentials with various
trap aspect ratios $\alpha = 1,2.5,5,7.5,10$. 
The chemical potentials are adjusted to fix the total particle number 
$N = N_\uparrow + N_\downarrow \simeq 210$ with the polarization 
$P = (N_\uparrow - N_\downarrow)/N$.
The particle density is below quarter filling at all lattice sites. 
The trap strength is given as $V_0=0.02\alpha^{2/3}$
to keep the volume constant.
The lattice size varies accordingly with $\alpha$:
$(L_x,L_y,L_z) = (24,24,24)$ for $\alpha=1$,  
$(16,16,40)$ for $\alpha=2.5$,
$(14,14,70)$ for $\alpha=5$, 
$(14,14,80)$ for $\alpha=7.5$,
and $(14,14,100)$ for $\alpha=10$. 
The lattice size becomes a bottleneck in the real-space DMFT. 
The computation took  $\sim 10^9$  CPU seconds in supercomputers. 

\begin{figure}
\includegraphics[width=0.47\textwidth]{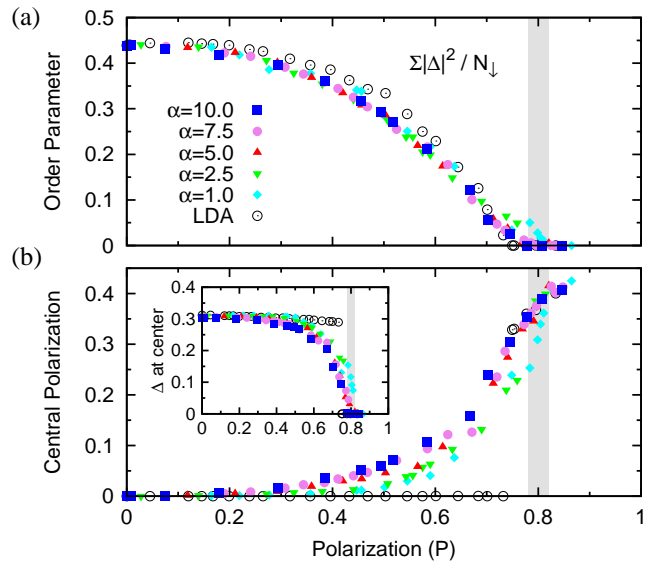} 
\caption{
Transition to the normal phase with increasing polarization $P$ 
examined for various trap aspect ratios $\alpha$.    
(a) Superfluid order parameter $\sum_i |\Delta_i|^2/N_\downarrow$ vanishes  
above $P_c \simeq 0.8$, regardless of $\alpha$.    
(b) Central polarization 
$(n_\uparrow - n_\downarrow)/(n_\uparrow + n_\downarrow)$ of the site 
at the cloud center becomes finite far below $P_c$, implying
the presence of a polarized superfluid core with 
finite $\delta n$ and $\Delta$ as indicated in the inset.   
The conventional DMFT data with LDA are given for comparison.
}
\label{fig2}
\end{figure}

Figure~\ref{fig1} shows how the distribution of particles and 
their superfluid characteristics change with polarization $P$ 
in an elongated trap with aspect ratio $\alpha=7.5$. 
The density difference 
$\delta n_i \equiv  n_{i\uparrow} - n_{i\downarrow}$ and 
the pair potential $\Delta_i \equiv 
\langle c^\dagger_{i\uparrow} c^\dagger_{i\downarrow} \rangle$ 
characterize the superfluid structure: 
for small $P$'s, the particles at the trap center are fully paired superfluid 
(dark gray, $\delta n=0$, finite $\Delta$) and surrounded by unpaired particles 
(light gray, $\delta n\neq 0$). 
The small bumps in $\Delta$ are found along the axial ($z$) direction, 
which can be interpreted as an analog of the proximity effect at 
a superconductor-ferromagnet interface~\cite{Buzdin2005}. 
As $P$ increases further, we find that two features become noticeable 
in $\Delta$. For $P \gtrsim 0.7$, an oscillatory structure 
in $\Delta$ emerges across the trap center, indicating that FFLO-type 
states may exist in this elongated system.
When $P$ reaches $0.8$, a transition to the normal phase occurs,
as indicated by the complete suppression of $\Delta$.

\begin{figure*}
\includegraphics[angle=-90,width=0.97\textwidth]{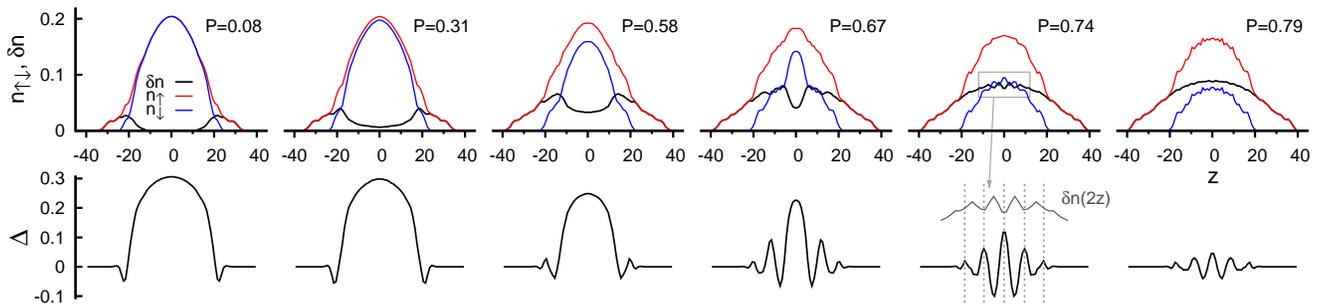} 
\caption{
Evolution of the cloud structure with increasing polarization $P$.  
Axial density profiles $n$ and pair potential $\Delta$ are shown
along the $z$-axis. The plots are selected 
for the trap aspect ratio $\alpha=7.5$. 
The coexistence of the finite dip in the density difference 
$\delta n \equiv n_\uparrow - n_\downarrow$ and 
a finite $\Delta$ at the central region characterizes
the polarized superfluid core.  
The central dip in $\delta n$ disappears at high $P=0.74$, replaced 
by the FFLO-type oscillations in $\Delta(z)$ spreading across the trap center.  
}
\label{fig3}
\end{figure*}

We study the transition to the normal phase systematically 
with the order parameter, $\eta \equiv \sum_i |\Delta_i|^2 / N_\downarrow$, 
averaged over the whole system. In Fig.~\ref{fig2}(a),
it turns out that the behavior of $\eta(P)$ and the
critical polarization $P_c \simeq 0.8$ is \textit{insensitive} 
to the trap aspect ratios that we have examined.
Moreover, we find that our $P_c$ agrees with $P_c \sim 0.78$ found
in the experiments \cite{Shin2006} and \cite{Nascimbene2009},
and the quantum Monte Carlo (QMC) results
in continuum with LDA~\cite{Lobo2006,Recati2008}.
The order parameter is comparable to 
the condensate fraction measured in \cite{Shin2006}.

On the other hand, we find that pairing in the superfluid core 
in our systems is unconventional. Figure~\ref{fig2}(b) indicates 
the coexistence of finite polarization and order parameter at the center.  
This characterizes the polarized superfluid core and is apparent 
at intermediate or even at low $P$'s for large $\alpha$. 
In contrast to the order parameter, the behavior of 
the central polarization depends on $\alpha$.
The LDA calculations, based on the conventional DMFT for homogeneous 
lattices at $T=0$, show a first-order transition from the fully paired superfluid phase 
to the partially paired normal phase without the intermediate polarized 
superfluid phase. However, the LDA cannot consider the inhomogeneous 
exotic phase observed in Fig.~\ref{fig1}, and it ignores 
the interface effects that may contribute to the polarized superfluidity.
While the potential role of finite-size effects should not be ruled out, 
we have tested also $N \simeq 320$ and found the same features.
 
Let us discuss in detail how the cloud structure 
and the pairing evolve as $P$ increases. 
Figure~\ref{fig3} presents the axial profiles of 
densities $n_{\uparrow,\downarrow}$,
their difference $\delta n$, and the pair potential
$\Delta$ along the $z$-axis. 
The data with $\alpha=7.5$ are shown as a representative example. 
The fully polarized edges are well separated from the superfluid core
with the partially polarized intermediate regions, 
while the central region shows nonuniform $\Delta$ 
that changes with $P$.  
Our LDA data (not shown) gives  the three-shell structure:
the fully paired superfluid core, the partially polarized normal state shell, 
and the fully polarized edges. This perfectly agrees with those from
the QMC with LDA~\cite{Recati2008}. While, at low $P$'s, 
our real-space DMFT presents a similar shell structure, 
the density profiles indicate the exotic superfluid phase at high $P$'s
that is not accessible by the LDA with conventional DMFT.

At very low polarization $P=0.08$,  the core is fully paired, 
and small oscillations of $\Delta$ appear in the partially polarized 
intermediate shoulders of $\delta n$. As $P$ increases, 
polarization becomes finite at center, leading to 
the wide dip area with finite $\delta n$, indicating 
the polarized superfluid core. 
The dip region gets narrower with increasing $P$ while
the shoulders with small $\Delta$ oscillations get wider. 

When $P$ increases further, the dip in $\delta n$ finally disappears,
and the FFLO-type oscillations in $\Delta$ spread across the trap 
center, as plotted for $P=0.74$ in Fig.~\ref{fig3}. 
As $P$ approaches $P_c$, the amplitude of $\Delta$ becomes smaller 
and finally vanishes at $P_c$. 
Along with the oscillating $\Delta$, we also find the corresponding 
oscillations in $\delta n$ with a half period of $\Delta$.
This is consistent with the prediction of the FFLO-type phase  
in one dimensional (1D) systems~\cite{Tezuka2008}. However, 
our systems still show 3D features: at low $P$'s,  
density profiles are similar to the LDA data in 3D while fully paired 
edges together with the FFLO phase at the center are expected
in the strictly 1D system~\cite{Liao2010,Tezuka2008}.

\begin{figure}[b]
\includegraphics[width=0.45\textwidth]{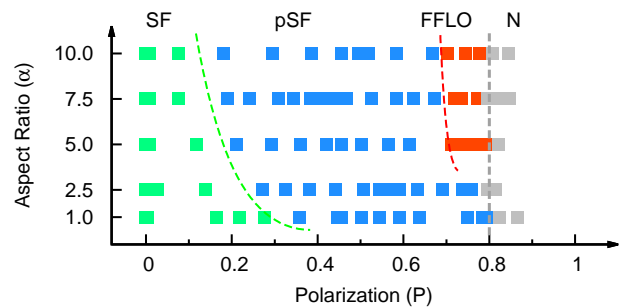} 
\caption{
Trap aspect ratio $\alpha$ and polarization $P$ 
dependence of the phase of the central region 
of the trap.  Fully paired superfluid (SF) is found at low $P$'s and 
continuously evolves into the polarized superfluid (pSF). 
The pSF is specified with the central polarization 
larger than $0.005$.  At higher $P$'s, the FFLO-type phase 
is identified for $\alpha \ge 5.0$.
The dashed lines between the phases are drawn
 for guidance.  
}
\label{fig4}
\end{figure}

We find that these FFLO-type oscillations emerging across the trap center, 
rather than residing just at the edges, 
are found only in highly elongated traps with the trap aspect ratios 
$\alpha \ge 5.0$ among the examined values of $\alpha$'s. 
The dependence of the core phase on $\alpha$  
is summarized in Fig.~\ref{fig4}.
For $\alpha=1.0$ and $2.5$, we have found the oscillations of
$\Delta$ reside only at the edges for all $P < P_c$.

\begin{figure}
\includegraphics[width=0.47\textwidth]{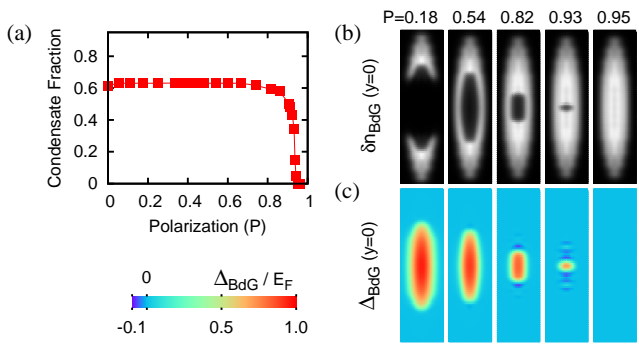} 
\caption{
BdG calculations at unitarity in the elongated trap 
with $\alpha=10$. 
(a) The condensate fraction
is plotted as a function of polarization $P$. 
(b) The density difference $\delta n_\mathrm{BdG}$ and (c) the pair potential 
$\Delta_\mathrm{BdG}$ (normalized by the Fermi energy $E_{F}$)
in the $y=0$ plane are shown for selected $P$'s.
The majority population $N_\uparrow$ is fixed at $200$ 
for a given $P$.
}
\label{fig5}
\end{figure}

The diagram in Fig.~\ref{fig4} separates 
the phases of the central region of the traps into 
fully paired superfluid (SF), polarized superfluid (pSF), 
FFLO, and normal (N). The FFLO-type phase can be 
signaled by the disappearance of the central dip in $\delta n$, 
as shown in Fig.~\ref{fig3}.
The boundary between SF and pSF is rather unclear because 
$\delta n$ becomes finite gradually with increasing $P$. 
However, our calculations show a tendency 
for the central polarization to grow faster as the trap gets more 
elongated.

The nature of the polarized superfluidity found in our 
real-space DMFT calculations at $T=0$ can be very subtle. 
It has not been clear yet how stable pSF is at $T=0$, especially in 
a trap. The BP state is in general energetically 
unstable toward the phase separation at a weak coupling at $T=0$
\cite{Bedaque2003,Koponen2007}. 
For a strong coupling, an earlier DMFT calculation in half-filled 
homogeneous cubic lattices found that the pSF may be 
stable down to a very low temperature~\cite{Dao2008},
while other calculations suggested that the density difference 
may decay exponentially with decreasing temperature in quarter-filled 
infinite dimensional lattices~\cite{Koga2010a}.  
In our LDA calculations, we have observed a sharp SF-N transition
without an intermediate pSF phase.
The stability issue can be more complex in our trapped 
asymmetric systems. We have found
nonstandard (pSF)-FFLO-N type structures in the elongated traps,
which cannot be explained with LDA. 

Our real-space DMFT inspires the question: how different are the results 
qualitatively from those given by the BdG \textit{static} mean-field theory,
incorporating a trapping potential~\cite{Jensen2007}? 
The DMFT is based on a discrete lattice system, 
nevertheless we choose to compare it to a BdG approach without a lattice 
(the density of our lattice fermions is well below the quarter filling).
In Fig.~\ref{fig5}, we observe very high $P_c$ ($\sim 0.95$) and 
the strong deformation of the core shape at high $P$'s. 
These features are in clear contrast with those found in our real-space 
DMFT. The DMFT accurately describes local quantum fluctuations 
which are missing in the BdG theory that does not properly include 
interactions in the normal state.
Such effects cause the large deviation in $P_c$ 
in the strongly interacting regime.     

The explicit test for the trap aspect ratio sheds light on the comparison
between the experiments \cite{Zwierlein2006,Shin2006} and \cite{Partridge2006}. 
Despite the difference between dilute gases and low-filling lattice 
fermions considered here, the critical polarization and the three-shell 
density profiles that we have found out are in good agreement
with \cite{Zwierlein2006,Shin2006} (and \cite{Nascimbene2009}). 
While the very large aspect ratio $\alpha \sim 45$ used
in  \cite{Partridge2006} is not accessible because of
computational limitations, we have not found any signatures of 
changing density profiles when increasing the trap aspect ratio.

The exotic superfluid phases found in our study suggest that high
polarizations $P \lesssim P_c$ can be an interesting area for future
experiments with spin-imbalanced Fermi gases. In our calculations
at $T=0$, we have found that the FFLO-like phase emerges at high
polarizations close to $P_c$. The amplitude of the oscillating
pair potential across the trap center is expected to be still
significantly large, which implies that it may be experimentally
accessible at low but finite temperatures. In addition, the density
and pair potential profiles that we have found in elongated traps
reveal the 3D characteristics which are essentially
different from the profiles predicted in strictly 1D.
This emphasizes the possibility for the experimental observation 
of the FFLO phase in 3D systems.

\begin{acknowledgments}
This work was supported by the Academy of Finland
(Projects No. 213362, No. 217043, No. 217045, No. 210953, No. 135000, and No. 139514)
and EuroQUAM/FerMix, and conducted as a part of
a EURYI scheme grant (see www.esf.org/euryi). 
Computing resources were provided by CSC -- 
the Finnish IT Centre for Science and the Triton cluster at the Aalto University.
\end{acknowledgments}

\end{document}